\documentclass[11pt,english]{article}
\usepackage[T1]{fontenc}
\usepackage{a4wide}
\usepackage{babel}

\makeatletter

\providecommand{\LyX}{L\kern-.1667em\lower.25em\hbox{Y}\kern-.125emX\@}

\usepackage{verbatim}

\makeatother

\begin{document}

\begin{flushright}CBPF-NF-028/01\\hep-th/0104171\\April 2001 

\end{flushright} 

\vspace{1cm}

\begin{center}

\textbf{\LARGE BPS String Solutions in Non-Abelian Yang-Mills Theories and Confinement}{\LARGE \par}

\vspace{1cm}

{\large {\bf Marco A. C. Kneipp\( ^{(a)(b)} \)}}\footnote{e-mail: {\tt kneipp@cbpf.br}} {\large {\bf and Patrick Brockill\( ^{(a)} \)}}\footnote{e-mail: {\tt brockill@cbpf.br}}

\vspace{0.3cm}

\end{center}

\begin{center}{\em \( ^{(a)} \)Centro Brasileiro de Pesquisas F\' \i sicas
(CBPF), \\ 

Coordenação de Teoria de Campos e Partículas (CCP), \\ Rua Dr. Xavier Sigaud,
150 \\

22290-180 Rio de Janeiro, Brazil. \\
\medskip{}

\( ^{(b)} \)Universidade do Estado do Rio de Janeiro(UERJ),\\

Depto. de Física Teórica,\\

Rua São Francisco Xavier, 524\\

20550-013 Rio de Janeiro, Brazil. \footnote{%
Address as of June 1, 2001.
}}

\end{center}

\begin{abstract}
Starting from the bosonic part of N=2 Super QCD with a ``Seiberg-Witten''
\( N=2 \) soft breaking mass term, we obtain string BPS conditions for arbitrary
semi-simple gauge groups. We show that the vacuum structure is compatible with
a symmetry breaking scheme which allows the existence of \( Z_{k} \)-strings
and which has \( \textrm{Spin}(10)\rightarrow SU(5)\times Z_{2} \) as a particular
case. We obtain BPS \( Z_{k} \)-string solutions and show that they satisfy
the same first order differential equations as the BPS string for the \( U(1) \)
case. We also show that the string tension is constant, which may cause a confining
potential between monopoles increasing linearly with their distance.

\vfill PACS numbers: 11.27.+d, 11.15.-q, 02.20.-a 
\end{abstract}
\newpage

\section{Introduction}

String (and vortex) solutions\cite{NO73} may have many important applications
such as their possible relevance for quark confinement \cite{MandeltHooft}\cite{SeibergWitten1},
for galaxy formation \cite{ZeldoVilen}\cite{KibbleLazaShafi} and for superconductors
\cite{Abrikosov}. These solutions may also be relevant for a field theory formalism
for the ``fundamental'' string or the D-strings. Non-BPS string solutions
in non-Abelian theories were first analysed in \cite{HK85}\cite{topchar},
for the particular case of \( SO(10) \). There are various motivations for
looking for BPS solutions. Firstly because they appear naturally in supersymmetric
theories, often in connection with dualities. Secondly, because they satisfy
first order differential equations which are easier to solve than the second
order equations coming from the equations of motion. And finally because BPS
(or almost-BPS) strings may be relevant for confinement\cite{SeibergWitten1}
(for a recent review see\cite{Yung00}).

The BPS solutions for monopoles in Yang-Mills-Higgs are known for an arbitrary
semi-simple gauge group broken by a scalar in the adjoint representation\cite{Bais78}.
However for strings (and vortices), the BPS solutions are only known for \( U(1) \)
Yang-Mills-Higgs theories broken to a \( Z \) exact symmetry group \cite{Bog(76)}
(see a review \cite{GM(86)}) and for some other particular examples (\cite{HK95}\cite{AchucarroVachapati}
and references therein). Our aim is thus to obtain the BPS string solutions
in a Yang-Mills-Higgs theory with an arbitrary semi-simple gauge group broken
to a non-Abelian residual group. 

In the paper of Seiberg-Witten \cite{SeibergWitten1}, the authors consider
a \( SU(2) \) \( N=2 \) super Yang-Mills theory, and associated to the point
in the moduli space where the monopole becomes massless they obtained an effective
\( U(1) \) \( N=2 \) super QED with an \( N=2 \) mass breaking term. In this
effective theory, the \( U(1) \) is broken to a \( Z \) group, the theory
develops an Abelian string solution and as it happens Abelian confinement occurs.
After this work, many interesting papers appeared \cite{varios} analysing various
related issues, considering either \( U(1) \) or \( U(1)^{N-1} \) effective
theories broken to discrete groups. Since we are considering a non-Abelian generalization
of Seiberg-Witten effective theory with a non-Abelian unbroken group, our BPS
string solution may have some relevance for non-Abelian confinement. More precisely,
in our theory the strings are associated to elements of a \( Z_{k} \) group,
rather than the \( Z \) group, and the breaking of gauge symmetry by a scalar
in the adjoint allows monopole solutions to arise belonging to representations
of (the dual) non-Abelian unbroken symmetry\cite{HolloDoFraMACK}, rather than
\( U(1) \) singlets. 

Keeping in mind that our results can be specifically applied to the symmetry
breaking scheme \( \textrm{Spin}(10)\stackrel{126}{\rightarrow }SU(5)\times Z_{2} \),
our BPS string solution may also have some applications as a cosmic string. 

We begin by obtaining, in section 2, the string BPS conditions considering the
bosonic part of \( N=2 \) super QCD with one flavour and with an \( N=2 \)
breaking mass term for the scalar in the vector multiplet, similar to the case
considered by Seiberg-Witten\cite{SeibergWitten1}. Then, in section 3 we show
that the vacuum structure is compatible with a symmetry breaking scheme considered
by Olive and Turok \cite{OT82}, which allows the existence of \( Z_{k} \)-strings
and which has \( \textrm{Spin}(10)\stackrel{126}{\rightarrow }SU(5)\times Z_{2} \)
as a particular case. In section 4, we consider a \( Z_{k} \)-string ansatz
and obtain the first order differential equations which are exactly the same
as the ones for the BPS string in the \( U(1) \) theory. From this ansatz we
obtain that the string tension is constant. This may ensure a confining potential
between monopoles increasing linearly with their distance.

\section{BPS String Conditions in Non-Abelian Yang-Mills-Higgs Theories.}

Let us consider the Lagrangian in 3+1 dimensions
\begin{equation}
\label{7.1}
L=\mbox {Tr}\left\{ -\frac{1}{4}G^{\mu \nu }G_{\mu \nu }+\frac{1}{2}D_{\mu }S^{\dagger }D^{\mu }S\right\} +\frac{1}{2}D_{\mu }\phi ^{\dagger }D^{\mu }\phi -V(S,\phi )\, .
\end{equation}
 with an arbitrary semi-simple gauge group, where \( S \) is a complex scalar
field in the adjoint representation and \( \phi  \) is another complex scalar
whose representation we shall specify below. As in the \( U(1) \) theory (for
a review see \cite{GM(86)}), let us consider a static field configuration with
cylindrical symmetry not depending on \( x^{3} \) and the only non-vanishing
component of \( G_{\mu \nu } \) being \( G_{12}\equiv -B \). The string tension
is then 
\begin{eqnarray}
T & = & \int d^{2}x\left\{ \frac{1}{2}\mbox {Tr}\left[ B^{2}+\left| D_{\mu }S\right| ^{2}\right] +\frac{1}{2}\left| D_{\mu }\phi \right| ^{2}+V(S,\phi )\right\} \nonumber \\
 & \geq  & \int d^{2}x\left\{ \frac{1}{2}\mbox {Tr}\left[ B^{2}+\left| D_{1}S\right| ^{2}+\left| D_{2}S\right| ^{2}\right] +\frac{1}{2}\left| D_{1}\phi \right| ^{2}+\frac{1}{2}\left| D_{2}\phi \right| ^{2}+V(S,\phi )\right\} \label{7.2} 
\end{eqnarray}
 Let us denote by \( \rho  \) the distance from the string axis. In order for
\( T \) to be finite, the field must tend to vacua configurations at \( \rho \rightarrow \infty  \)
, satisfying the conditions 
\begin{eqnarray}
D_{\mu }S=D_{\mu }\phi  & = & O(1/\rho ^{2})\, ,\nonumber \\
V(S,\phi ) & = & O(1/\rho ^{3})\, ,\label{7.3} \\
B & = & O(1/\rho ^{2})\, .\nonumber 
\end{eqnarray}
 Let \( D_{\pm }=D_{1}\pm iD_{2} \). Using the identity
\[
\left[ D_{\pm }\phi \right] ^{\dagger }\left[ D_{\pm }\phi \right] -\left| D_{1}\phi \right| ^{2}-\left| D_{2}\phi \right| ^{2}=\pm \left[ i\epsilon _{ij}\partial _{i}\left( \phi ^{\dagger }D_{j}\phi \right) +e\phi ^{\dagger }G_{12}\phi \right] \, ,\]
 and the fact that
\[
\int d^{2}x\epsilon _{ij}\partial _{i}\left( \phi ^{\dagger }D_{j}\phi \right) =0\, ,\]
(which follows from the above boundary conditions) and similar results for the
scalar \( S \), it results that
\begin{eqnarray*}
T & = & \int d^{2}x\left\{ \mbox {Tr}\left[ \frac{1}{2}B^{2}+\frac{1}{2}\left| D_{\mp }S\right| ^{2}\mp \frac{e}{2}S^{\dagger }\left[ B,S\right] \right] +\right. \\
 &  & \left. +\frac{1}{2}\left| D_{\pm }\phi \right| ^{2}\pm \frac{e}{2}\left( \phi ^{\dagger }B\phi \right) +V(S,\phi )\right\} \\
 & \geq  & \int d^{2}x\left\{ \frac{1}{2}B_{a}^{2}\pm \frac{e}{2}\left( S_{b}^{*}if_{bca}S_{c}+\phi ^{\dagger }T_{a}\phi \right) B_{a}+V(S,\phi )\right\} \, .
\end{eqnarray*}
Note that we used the above identities with opposite signs for the fields \( \phi  \)
and \( S \), in order to make contact with the supersymmetric Lagrangian, as
will become clear below. Let
\begin{equation}
\label{7.3a}
Y_{a}=\frac{e}{2}\left( S^{*}_{b}if_{bca}S_{c}+\phi ^{\dagger }T_{a}\phi \right) +X_{a}
\end{equation}
 where
\[
X_{a}=-\frac{me}{2}\left( \frac{S_{a}+S_{a}^{*}}{2}\right) \, .\]
 Then \( T \) can be written as 
\begin{eqnarray}
T & \geq  & \int d^{2}x\left\{ \frac{1}{2}\left[ B_{a}\pm Y_{a}\right] ^{2}\mp X_{a}B_{a}-\frac{1}{2}Y^{2}_{b}+V(S,\phi )\right\} \nonumber \\
 & \geq  & \int d^{2}x\left\{ \mp X_{a}B_{a}-\frac{1}{2}Y_{a}^{2}+V(S,\phi )\right\} \, .\nonumber \label{7.4} 
\end{eqnarray}
If \( V(S,\phi )-Y_{a}^{2}/2\geq 0 \), then
\begin{equation}
\label{7.5}
T\geq \int d^{2}x\left\{ \mp X_{a}B_{a}\right\} 
\end{equation}
and the bound is saturated if and only if
\begin{eqnarray}
D_{0}\phi  & = & D_{3}\phi =D_{0}S=D_{3}S=0\label{7.6aa} \\
D_{\pm }\phi  & = & 0\, ,\label{7.6a} \\
D_{\mp }S & = & 0\, ,\label{7.6c} \\
B_{a}\pm Y_{a} & = & 0\, ,\label{7.6d} \\
V(S,\phi )-\frac{1}{2}Y_{a}^{2} & = & 0\, ,\label{7.6e} 
\end{eqnarray}
which are BPS equations for the string. The first conditions (\ref{7.6aa})
imply that \( W_{0}=0=W_{3} \).

We shall consider
\begin{equation}
\label{7.7}
V(S,\phi )=\frac{1}{2}\left( Y_{a}^{2}+F^{\dagger }F\right) \, \, \, \, ,\, \, \, \, \, \, F\equiv e\left( S^{\dagger }-\frac{\mu }{e}\right) \phi \, .
\end{equation}
 Then the BPS condition (\ref{7.6e}) implies that \( F=0 \). When \( m=0 \),
this potential coincides with the one for the bosonic part of \( N=2 \) Super
QCD with one flavour (\ref{4.1}) (see appendix A) if the scalar \( \phi _{2}=0 \)
of the hypermultiplet vanishes. The case \( m\neq 0 \) clearly breaks \( N=2 \)
supersymmetry since it gives a bare mass to \( S_{a} \). This is akin to the
situation considered by Seiberg and Witten \cite{SeibergWitten1} where the
authors obtained confinement by introducing a bare mass to the scalar in the
vector supermultiplet. This \( X_{a} \) term is important in order to change
the vacuum structure of the theory.

The equations of motion that follow from our Lagrangian are
\begin{eqnarray*}
\left( D_{\mu }G^{\mu \nu }\right) _{a}-\frac{ie}{2}\left( \phi ^{\dagger }T_{a}D^{\nu }\phi -D^{\nu }\phi ^{\dagger }T_{a}\phi -S_{b}^{*}if_{abc}D^{\nu }S_{c}+D^{\nu }S^{*}_{b}if_{abc}S_{c}\right)  & = & 0\, ,\\
D_{\mu }D^{\mu }\phi _{k}+eY_{a}T^{a}_{kl}\phi _{l}-e^{2}\left[ \left( S-\frac{\mu }{e}\right) \left( S^{\dagger }-\frac{\mu }{e}\right) \phi \right] _{k} & = & 0\, ,\\
D_{\mu }D^{\mu }S_{d}+eY_{a}\left( if_{adc}S_{c}-\frac{m}{2}\delta _{ad}\right) -e^{2}\phi ^{\dagger }\left( S-\frac{\mu }{e}\right) T_{d}\phi  & = & 0\, .
\end{eqnarray*}
 Let us check if the BPS equations for the string are consistent with them.
Acting with the covariant derivative \( D_{i} \), \( i=1,2 \), on (\ref{7.6d})
and using the other BPS conditions it results that
\[
D_{\mu }G^{\mu \nu }_{a}+\frac{ie}{2}\left[ D^{\nu }\phi ^{\dagger }T_{d}\phi -\phi ^{\dagger }T_{d}D^{\nu }\phi -\left( D^{\nu }S\right) ^{*}_{b}if_{dbc}S_{c}+S_{b}^{*}if_{dbc}D^{\nu }S_{c}-\frac{m}{2}\left( D^{\nu }S_{d}-D^{\nu }S^{*}_{d}\right) \right] =0\]
This relation is consistent with the first equation of motion if \( m=0 \).
Similarly, from the BPS conditions we obtain
\begin{eqnarray*}
0 & = & D_{\mp }D_{\pm }\phi -e\left( S-\frac{\mu }{e}\right) F\\
 & = & -D_{\mu }D^{\mu }\phi \mp eG_{12}\phi -e\left( S-\frac{\mu }{e}\right) F\\
 & = & -D_{\mu }D^{\mu }\phi -eY_{a}T_{a}\phi -e\left( S-\frac{\mu }{e}\right) F
\end{eqnarray*}
and 
\begin{eqnarray*}
0 & = & D_{\pm }D_{\mp }S_{d}-eF^{\dagger }T_{d}\phi \\
 & = & -D_{\mu }D^{\mu }S_{d}\pm e\left( G_{12}S\right) _{d}-eF^{\dagger }T_{d}\phi \\
 & = & -D_{\mu }D^{\mu }S_{d}-ieY_{a}f_{adb}S_{b}-eF^{\dagger }T_{d}\phi 
\end{eqnarray*}
Once again, this last relation is consistent with the equations of motion only
when \( m=0 \). However this condition must be understood in the limiting case
\( m\rightarrow 0 \) as we shall discuss in the next section. Therefore it
is only in this limit that we can have BPS strings satisfying (\ref{7.6aa})-(\ref{7.7}).
One can check that 1/4 of the \( N=2 \) supersymmetry transformations (\ref{a1})
vanish for field configurations satisfying the BPS conditions (\ref{7.6aa})-(\ref{7.7})
in the limit \( m\rightarrow 0 \).

\section{Vacua solutions.}

The total energy for this theory is non-negative and it vanishes (vacuum) if
and only if 
\begin{eqnarray}
D_{\mu }\phi  & = & D_{\mu }S=G_{\mu \nu }=0\, ,\label{3.1} \\
V & = & 0\, \, \, \, \, \Leftrightarrow \, \, \, \, \, Y_{a}=F=0\nonumber 
\end{eqnarray}
in all spacetime. In order for the string to have finite tension \( T \), the
fields at \( \rho \rightarrow \infty  \) must tend to vacuum configurations.
Moreover, a necessary condition for the existence of string solutions is that
these vacua must break the gauge group \( G \) into \( G_{\phi } \) such that
\begin{equation}
\label{3.1a}
\Pi _{1}(G/G_{\phi })=Z_{k\, ,}
\end{equation}
which allows the existence of \( Z_{k} \) strings. 

Let us consider \( H_{i},\, E_{\alpha } \) to be generators of a Lie algebra
in the Cartan-Weyl basis, with \( H_{i}^{\dagger }=H_{i} \) and \( E_{\alpha }^{\dagger }=E_{-\alpha } \),
\( \textrm{Tr}(H_{i}H_{j})=\delta _{ij} \), \( \textrm{Tr}(E_{\alpha }E_{-\beta })=2\delta _{\alpha \beta }/\alpha ^{2} \)
and satisfying the commutation relations
\begin{eqnarray*}
\left[ H_{i},E_{\alpha }\right]  & = & \alpha ^{i}E_{\alpha }\, ,\\
\left[ E_{\alpha },E_{-\alpha }\right]  & = & \alpha ^{\textrm{v}}\cdot H\, ,\, \, \, \, \, \, \, \, \, \, \alpha ^{\textrm{v}}\equiv \frac{2\alpha }{\alpha ^{2}}\, .
\end{eqnarray*}
Moreover \( H_{i}|\lambda _{a}>=\lambda ^{i}_{a}|\lambda _{a}> \). A symmetry
breaking satisfying (\ref{3.1a}) can be realized in the following way\cite{OT82}:
let \( \lambda _{\phi } \) be an arbitrary fundamental weight and let \( S^{\textrm{vac}} \)
and \( \phi ^{\textrm{vac}} \) be two scalars in the vacuum configuration.
As is well known, a scalar in the adjoint representation of the form \( S^{\textrm{vac}}\propto \lambda _{\phi }\cdot H \)
breaks the gauge group into 
\[
G\rightarrow G_{S}=U(1)\times K/Z_{l}\, ,\]
where \( K \) is the subgroup of \( G \) associated to the algebra whose Dynkin
diagram is given by removing the dot corresponding to \( \alpha _{\phi } \)
from that of \( G \), \( U(1) \) is generated by \( \lambda _{\phi }\cdot H \)
and \( Z_{l} \) is a subgroup of \( U(1) \) and \( K \) and is generated
by 
\begin{equation}
\label{3.1aa}
v_{0}=\exp (2\pi iz\lambda ^{\textrm{v}}_{\phi }\cdot H)\, ,\, \, \, \, \, z\equiv \frac{\left| Z(G)\right| }{\left| Z(K)\right| }\, ,
\end{equation}
 where \( |Z(G)| \) is the order of the center of \( G \) and \( |Z(K)| \)
is the order of the center of \( K \). This symmetry breaking pattern allows
the existence of monopoles. If the theory has another scalar \( \phi ^{\textrm{vac}}\propto |k\lambda _{\phi }> \),
\( k \) being an integer, the gauge group \( G \) is further broken into\cite{OT82}
\[
G\rightarrow G_{\phi }=Z_{kl}\times K/Z_{l}\subset G_{S}\]
 where \( Z_{kl} \) is generated by \( v_{0}^{1/k} \) and \( K \) is as before.
Then, \( \Pi _{1}(G/G_{\phi })=Z_{k} \). In order for \( \phi ^{\textrm{vac}}\propto |k\lambda _{\phi }> \)
we may consider that \( \phi  \) belongs to the irrep \( R_{k\lambda _{\phi }} \)
with \( k\lambda _{\phi } \) as highest weight.\footnote{%
If \( k=2 \), it can also be interesting to consider \( \phi  \) belonging
to the symmetric part of the tensor product of two fundamental representations
with highest weight \( \lambda _{\phi } \), \( [R_{\lambda _{\phi }}\times R_{\lambda _{\phi }}]_{S}\equiv R^{\mbox {sym}}_{2\lambda _{\phi }}\supset R_{2\lambda _{\phi }} \).
(For \( SU(N) \), \( R^{\mbox {sym}}_{2\lambda _{N-1}}=R_{2\lambda _{N-1}} \).
For \( SO(10) \), \( R^{\mbox {sym}}_{2\lambda _{5}}=126\oplus 10 \) and \( R_{2\lambda _{5}}=126 \)
). A physical motivation to consider \( \phi \in  \)\( R^{\mbox {sym}}_{2\lambda _{\phi }} \)
is because it allows a Yukawa coupling with two spinors in the fundamental representation
\( R_{\lambda _{\phi }} \), and this term gives rise to the mass term for the
spinors when \( \phi  \) has a non-trivial expectation value. In this case
one could also consider \( \phi  \) as a difermion condensate as in the BCS
theory.
} The symmetry breaking scheme \( \textrm{Spin}(10)\stackrel{126}{\rightarrow }SU(5)\times Z_{2} \)
considered by Kibble et al\cite{KibbleLazaShafi} for the cosmic string corresponds
to a particular case of this general result.

Let us show that the vacuum conditions (\ref{3.1}) admit solutions of this
form. Consider \( \phi ^{\textrm{vac}}=a|k\lambda _{\phi }> \) and \( S^{\textrm{vac}}=v\cdot H \)
where \( a\, ,\, v\in R \) and \( <k\lambda _{\phi }|k\lambda _{\phi }>=1 \).
The condition \( Y_{a}=0 \) is equivalent to
\begin{equation}
\label{8.1}
Y_{a}T_{a}=\frac{e}{2}\left[ \left( \phi ^{\dagger }T_{a}\phi \right) T_{a}+\left[ S^{\dagger },S\right] -m\left( \frac{S+S^{\dagger }}{2}\right) \right] =0\, .
\end{equation}
 Using that
\begin{eqnarray*}
(\phi _{i}^{*}T_{aij}\phi _{j})T_{a} & = & \textrm{Tr}\left( \phi \phi ^{\dagger }T_{a}\right) T_{a}=\phi \phi ^{\dagger }=\\
 & = & \left( \phi ^{\dagger }H_{i}\phi \right) H_{i}+\frac{\alpha ^{2}}{2}\left( \phi ^{\dagger }E_{\alpha }\phi \right) E_{-\alpha }\, ,
\end{eqnarray*}
 from (\ref{8.1}) follows that 
\[
v=\frac{ka^{2}}{m}\lambda _{\phi \, .}\]
On the other hand, from the condition \( F=0 \), it results that \( v\cdot \lambda _{\phi }=\mu /ke \),
which together with the previous relation leads to
\[
a^{2}=\frac{m\mu }{k^{2}e\lambda ^{2}_{\phi }}\, .\]
Then,
\begin{eqnarray}
\phi ^{\textrm{vac}} & = & a|k\lambda _{\phi }>\, ,\nonumber \\
mS^{\textrm{vac}} & = & ka^{2}\lambda _{\phi }\cdot H\, ,\label{3.5} \\
W_{\mu }^{\textrm{vac}} & = & 0\, ,\nonumber 
\end{eqnarray}
 is a solution of the vacuum conditions (\ref{3.1}) which satisfy \( \Pi _{1}(G/G_{\phi })=Z_{k} \). 

Expanding the fields around this vacuum (\( S=S^{q}+S^{\textrm{vac}}, \) \( W_{\mu }=W_{\mu }^{q}+W_{\mu }^{\textrm{vac}} \),
etc) and considering
\[
W_{\mu }^{q}=W^{\phi }_{\mu }H_{\alpha _{\phi }}+\sum _{i\neq \phi }W_{\mu }^{i}H_{\alpha _{i}}+\sum _{\alpha }W_{\mu }^{\alpha }E_{\alpha }\, ,\]
 where \( H_{\alpha }\equiv \alpha ^{\textrm{v}}\cdot H \), from the kinetic
terms of \( S \) and \( \phi  \) one finds the gauge particle mass terms
\[
\sum _{\alpha >0}\left( \lambda _{\phi }\cdot \alpha ^{\textrm{v}}\right) \left[ \frac{\left( \alpha \right) ^{4}e^{2}k^{2}a^{4}}{4m^{2}}+\frac{e^{2}a^{2}k}{2}\right] W^{\alpha \mu }W^{-\alpha }_{\mu }+\frac{e^{2}a^{2}k^{2}}{2}W^{\phi \mu }W^{\phi }_{\mu }\, .\]

As we mentioned before, the BPS conditions are compatible with the equations
of motion when \( m=0 \). However, if we do this, \( a=0 \) and there is no
symmetry breaking, which is necessary in order for string solutions to exist.
This result is very similar to what happens for the BPS monopole (see for instance
\cite{GoddardOlive}). In that case, one of the BPS conditions is \( V(\phi )=\lambda (\phi ^{2}-a^{2})^{2}/4=0 \),
which implies the vanishing of the coupling \( \lambda  \).(Note that for the
string and the monopole, \( X_{a} \) and \( V \) are terms which break \( N=2 \)
supersymmetry and which vanish for the BPS configurations.) However, that condition
must be understood in the Prasad-Sommerfield limiting case \( \lambda \rightarrow 0 \)
\cite{PrasadSommerfield} in order to retain the boundary condition \( |\phi |\rightarrow a \)
as \( r\rightarrow \infty  \), and to have symmetry breaking. In our case,
we have the same situation with a small difference: if one considers \( m\rightarrow 0 \),
then \( a\rightarrow 0 \). We can avoid this problem by allowing \( \mu \rightarrow \infty  \)
such that \( m\mu  \), or equivalently \( a \), remains constant, implying
that the field \( \phi  \) becomes infinitely heavy. The same happens for the
gauge fields \( W_{\mu }^{\alpha } \) in which \( \lambda _{\phi }\cdot \alpha ^{\textrm{v}}\neq 0 \).

It is important to mention that if we take \( m=0 \), (\ref{3.5}) is no longer
a vacuum solution, but it is possible to consider other vacuum solutions such
that \( \Pi _{1}(G/G_{\phi })\neq 0 \). However, in this case, we were not
able to construct a string ansatz satisfying the BPS conditions.

\section{BPS String Solutions.}

The string must tend at \( \rho \rightarrow \infty  \) to vacuum solutions
in any angular direction \( \varphi  \). Let us denote \( \phi (\varphi )=\phi (\varphi ,\rho \rightarrow \infty ), \)
\( S(\varphi )=S(\varphi ,\rho \rightarrow \infty ) \). Then, the vacuum conditions
(\ref{3.1}) imply that this asymptotic field configuration must be related
by gauge transformations from a vacuum configuration, which we shall consider
(\ref{3.5}), i.e.
\begin{eqnarray*}
W_{i}(\varphi ) & = & \frac{-1}{ie}\left( \partial _{i}g(\varphi )\right) g(\varphi )^{-1}\, \, \, \, \, \, \, \, \, i=1,2\, ,\\
\phi (\varphi ) & = & g(\varphi )\phi ^{\textrm{vac}}\, ,\\
S(\varphi ) & = & g(\varphi )S^{\textrm{vac}}g(\varphi )^{-1}\, ,
\end{eqnarray*}
for some \( g(\varphi )\in G \). Then, in order for the field configurations
to be single-valued, \( g(2\pi )g(0)\in G_{\phi } \). Without lost of generality
we shall consider \( g(0)=1 \). We shall also consider that \( G \) is simply
connected (which can always be done by going to the universal covering group).
Then, a necessary condition for the existence of strings is that \( g(2\pi ) \)
belongs to a non-connected component of \( G_{\phi } \)\cite{HK85}. Let \( g(\varphi )=\exp i\varphi M \).
Then, at \( \rho \rightarrow \infty  \) 
\begin{eqnarray}
\phi (\varphi ) & = & ae^{i\varphi M}|k\lambda _{\phi }>\, ,\nonumber \\
mS(\varphi ) & = & ka^{2}e^{iM\varphi }\lambda _{\phi }\cdot He^{-iM\varphi }\, ,\label{3.4} \\
W_{i}(\varphi ) & = & \frac{\epsilon _{ij}x^{j}}{e\rho ^{2}}M\, ,\, \, \, \, \, \, \, i,j=1,2\, .\nonumber 
\end{eqnarray}
A possible choice for \( M \) is
\[
M=\frac{n}{k}\frac{\lambda _{\phi }\cdot H}{\lambda _{\phi }^{2}}\, ,\]
with \( n \) being a non-vanishing integer defined modulo \( k \). From (\ref{3.4}),
it is direct to see that for this choice \( g(2\pi )\in G_{\phi } \). Indeed,
since\cite{GoddardOlive81} 
\[
\lambda ^{2}_{\phi }=\frac{1}{2}\alpha _{\phi }^{2}\frac{\left| Z(K)\right| }{\left| Z(G)\right| }\, ,\]
we see from (\ref{3.1aa}) that \( g(2\pi )=v_{0}^{n/k} \).

Let us consider the ansatz 
\begin{eqnarray}
\phi (\varphi ,\rho ) & = & f(\rho )e^{i\varphi M}a|k\lambda _{\phi }>\, ,\nonumber \\
mS(\varphi ,\rho ) & = & h(\rho )ka^{2}e^{i\varphi M}\lambda _{\phi }\cdot He^{-i\varphi M}\, ,\label{4.4} \\
W_{i}(\varphi ,\rho ) & = & g(\rho )M\frac{\epsilon _{ij}x^{j}}{e\rho ^{2}}\, \, \, \, \, \, \, \rightarrow \, \, \, \, \, \, \, B(\varphi ,\rho )=\frac{M}{e\rho }g'(\rho )\, ,\nonumber \\
W_{0}(\varphi ,\rho ) & = & W_{3}(\varphi ,\rho )=0\, ,
\end{eqnarray}
with the boundary conditions
\[
f(\infty )=g(\infty )=h(\infty )=1\, ,\]
in order to recover the configuration (\ref{3.4}) at \( \rho \rightarrow \infty  \)
and 
\[
f(0)=g(0)=0\]
in order to eliminate singularities at \( \rho =0 \).

Putting this ansatz in the BPS conditions (\ref{7.6a})-(\ref{7.6e}), from
the first order differential equations it results that:
\begin{eqnarray*}
h(\rho ) & = & \textrm{const }=1\\
f'(\rho ) & = & \pm \frac{n}{\rho }\left[ 1-g(\rho )\right] f(\rho )\\
g'(\rho ) & = & \mp \frac{e^{2}a^{2}\rho k^{2}\lambda _{\phi }^{2}}{2n}\left[ |f(\rho )|^{2}-1\right] 
\end{eqnarray*}
which are exactly the same differential equations which appear in the \( U(1) \)
case. These equations don't have analytic solutions, however their existence
has been proven and some of their properties have been analysed (see for instance
\cite{JaffeTaubes}). 

It is important to emphasize that the BPS conditions only hold when \( m\rightarrow 0 \)
and for \( m\neq 0 \) the string becomes non-BPS as has been already pointed
out in \cite{VaYung} for the \( G=U(1) \) case.

Using the ansatz (\ref{4.4}), it is straightforward to obtain the BPS bound
for the string tension (\ref{7.5}) 
\[
T=\pi a^{2}|n|\]
which once more coincides with the \( U(1) \) result. Since the tension is
constant, it may cause a confining potential between monopoles increasing linearly
with their distance, which is an interesting behavior since it may produce quark
confinement in a possible dual theory.

\section{Conclusions}

In this paper we showed the existence of BPS \( Z_{k} \)-string solutions for
arbitrary semi-simple gauge groups broken to non-Abelian groups. In order to
obtain these solutions we considered the bosonic part of \( N=2 \) SQCD with
one flavour and a \( N=2 \) breaking mass term. We showed that BPS conditions
are compatible with the equations of motion only if \( m\rightarrow 0 \). We
must also to take \( \mu \rightarrow \infty  \), with \( m\mu  \) fixed, in
order to allow gauge symmetry breaking, where \( m \) is the \( S \) bare
mass and \( \mu  \) is the \( \phi  \) bare mass. We found vacua solutions
compatible with the existence of string solutions and we were able to construct
these string solutions satisfying the BPS conditions. Since our theory is a
non-Abelian generalization of Seiberg-Witten effective theory, we hope that
our BPS string solution may have some relevance for non-Abelian confinement.
In particular, since in our theory the breaking of gauge symmetry by \( S \)
allows for monopole solutions belonging to representations of (the dual) non-Abelian
unbroken symmetry and the string solutions are associated to elements of a \( Z_{k} \)
group, we expect that monopole bound states with properties more similar to
the ones of quark bound states in QCD may appear in our theory. An indication
of the existence of these monopole bound states comes from the fact that in
our theory the BPS string tensions are constant which may give rise to a potential
between monopoles increasing linearly with their distance. It would be interesting
if one could find monopole bound solutions (in the classical theory) similar
to the breathers in sine-Gordon theory. 

\vskip 0.2 in \noindent \textbf{}\textbf{\large Acknowledgments} \textbf{}

\noindent We would like to thank M. Sarandy for collaboration in the early stages
of this work, as well as the kind hospitality of ``Grupo de Física Teórica''
of Universidade Católica de Petrópolis (UCP) where part of this work was conceived,
and in particular J.A. Helayël-Neto. M.A.C.K. also would like to thank N. Berkovits,
T. Kibble, D. Olive, R. Paunov and A. Schwimmer for many valuable conversations
and FAPERJ for financial support. P.B. would like to thank CAPES for financial
support.

\appendix

\section{{\large N=2 SQCD Potential}.}

Using Sohnius' conventions\cite{Sohnuis} and considering \( S=M+iN \), the
bosonic part for the potential of N=2 Super Yang-Mills with one hypermultiplet
can be written as
\begin{eqnarray}
V(S,\phi _{m}) & = & \frac{e^{2}}{8}\left\{ \left( S_{b}^{*}if_{bca}S_{c}\right) ^{2}+\left( \phi _{m}^{\dagger }\sigma ^{p}_{mn}T_{a}\phi _{n}-v_{p}\delta _{a0}\right) ^{2}+\frac{4\mu ^{2}}{e^{2}}\phi _{m}^{\dagger }\phi _{m}\right. \nonumber \\
 &  & \left. -\frac{4\mu }{e}\phi ^{\dagger }_{m}\left( S+S^{\dagger }\right) \phi _{m}+2\phi _{m}^{\dagger }\left\{ S^{\dagger },S\right\} \phi _{m}\right\} \label{2.2a} 
\end{eqnarray}
where \( \sigma ^{p} \) are the Pauli matrices and the terms \( v_{p}\delta _{a0} \)
are the Fayet-Iliopoulos that may exist associated to a possible \( U(1) \)
factor\footnote{%
The coupling constant for a possible \( U(1) \) factor is not necessarily the
same as the non-Abelian part, but for notational simplicity we shall consider
the same constant \( e \).
} with a generator we shall denote \( T_{0} \). This expression can be rewritten
as

\begin{equation}
\label{4.1}
V(S,\phi _{m})=\frac{1}{2}\left( \left( d_{a}^{1}\right) ^{2}+\left( d^{2}_{a}\right) ^{2}+\left( D_{a}\right) ^{2}+F_{m}^{\dagger }F_{m}\right) 
\end{equation}
where 
\begin{eqnarray}
D_{a} & = & \frac{e}{2}\left( S^{*}_{b}if_{bca}S_{c}\right) +d^{3}_{a}\, \, ,\label{4.2a} \\
d_{a}^{p} & = & \frac{e}{2}\left( \phi _{m}^{\dagger }\sigma ^{p}_{mn}T_{a}\phi _{n}-v_{p}\delta _{a0}\right) \, \, ,\, \, \, \, p=1,2,3\, \, ,\label{4.2b} \\
F_{1} & = & e\left( S^{\dagger }-\frac{\mu }{e}\right) \phi _{1}\, \, ,\label{4.2c} \\
F_{2} & = & e\left( S-\frac{\mu }{e}\right) \phi _{2}\, \, .\label{4.2d} 
\end{eqnarray}
From this expression it is easy to see that we recover (\ref{7.7}), for \( m\rightarrow 0 \),
when one puts \( \phi _{2}=0 \).

Let us denote by \( \psi  \) and \( \lambda ^{m},\, m=1,2 \), the pseudo-Majorana
spinors belonging to the vector supermultiplet and to the hypermultiplet respectively.
Their \( N=2 \) supersymmetry transformations are given by\cite{Sohnuis}
\begin{eqnarray}
\delta \lambda ^{m} & = & \frac{i}{2}G_{\mu \nu }\gamma ^{\mu \nu }\xi ^{m}-\gamma ^{\mu }D_{\mu }\left( M+\gamma _{5}N\right) \xi ^{m}-ie\left[ M,N\right] \gamma _{5}\xi ^{m}+i\xi ^{n}\sigma ^{p}_{nm}d^{p}\label{a1} \\
\delta \psi  & = & -\left[ i\gamma ^{\mu }D_{\mu }+e\left( M+\gamma _{5}N\right) -\mu \right] \xi ^{m}\phi _{m}\nonumber 
\end{eqnarray}
where \( \gamma ^{\mu \nu }\equiv i[\gamma ^{\mu },\gamma ^{\nu }]/2 \) and
where the \( \xi ^{m} \) are supersymmetry parameters.


\begin{thebibliography}{10}
\bibitem{NO73}H.B. Nielsen and P. Olesen, Nucl. Phys. \textbf{B61}(1973)45.
\bibitem{MandeltHooft}S. Mandelstam, Phys. Rep. \textbf{23C} (1976) 145; G. 'tHooft, in Proceed. of
Euro. Phys. Soc. 1975, ed A. Zichichi.
\bibitem{SeibergWitten1}N. Seiberg and E. Witten, Nucl. Phys. \textbf{B426} (1994)19.
\bibitem{ZeldoVilen}Ya.B. Zel'dovich, Mon. Not. R. Astron. Soc. \textbf{192} (1980) 663; A. Vilenkin,
Phys. Rev. Lett. \textbf{46} (1981)1169.
\bibitem{KibbleLazaShafi}T.W.B. Kibble, G. Lazarides and Q. Shafi, Phys. Rev. \textbf{D26} (1982) 435.
\bibitem{Abrikosov}A.A. Abrikosov, Sov. Phys. JETP \textbf{5} (1957) 1174.
\bibitem{HK85}M. Hindmarsh and T.W.B. Kibble, Phys. Rev. Lett. \textbf{55} (1985) 2398.
\bibitem{topchar}M. Aryal and A.E. Everett, Phys. Rev. \textbf{D35} (1987) 3105; 2398; C-P Ma,
Phys. Rev. \textbf{D48} (1993) 530.
\bibitem{Yung00}A. Yung, ``What do we learn about confinement from the Seiberg-Witten Theory'',
hep-th/0005088.
\bibitem{Bais78}F.A. Bais, Phys. Rev. D18 (1978) 1206.
\bibitem{Bog(76)}E.B. Bogomol'nyi, Sov. Jour. Nucl. Phys. \textbf{24} (1976) 449.
\bibitem{GM(86)}P. Goddard and P. Mansfield, Rep. Prog. Phys. \textbf{49} (1986) 725; P. Townsend
``Three Lectures on Supersymmetry and Extended Objects'' published in the
proceedings of the 13th GIFT Seminar on Theoretical Physics.
\bibitem{HK95}M.B. Hindmarsh and T.W.B. Kibble, Rept.Prog.Phys. \textbf{58} (1995) 477.
\bibitem{AchucarroVachapati}A. Achucarro and T. Vachapati, Phys.Rept.\textbf{327} (2000) 347.
\bibitem{OT82}D. Olive and N. Turok, Phys. Lett. \textbf{117B} (1982)193.
\bibitem{varios}M. Douglas and S.H. Shenker, Nucl.Phys. \textbf{B447} (1995) 271; Philip C.
Argyres and Michael R. Douglas, Nucl.Phys. \textbf{B448} (1995) 93; E. Witten,
Nucl.Phys. \textbf{B507} (1997) 658; A. Hanany, M. J. Strassler and A. Zaffaroni,
Nucl.Phys. \textbf{B513} (1998) 87; A. Yung, Nucl.Phys. \textbf{B562} (1999)
191; X. Hou, Phys.Rev. \textbf{D63} (2001) 045015; A. Gorsky, A. Vainshtein,
A. Yung, Nucl.Phys. \textbf{B584} (2000) 197; J. D. Edelstein, W. G. Fuertes,
J. Mas, J. M. Guilarte, Phys.Rev. \textbf{D62} (2000) 065008; A. Vainshtein
and A. Yung, hep-th/0012250; A Yung, hep-th/0103222.
\bibitem{HolloDoFraMACK}N. Dorey, C. Fraser, T. J. Hollowood and M.A.C. Kneipp, ``Non Abelian Duality
in N=4 Supersymmetric Gauge Theories'', hep-th/9512116. 
\bibitem{GoddardOlive}P. Goddard and D.I. Olive, Rep. Prog. Phys. \textbf{127} (1978) 1357.
\bibitem{PrasadSommerfield}M. K. Prasad and C.M. Sommerfield, Phys. Rev. Lett. 35 (1975) 445.
\bibitem{GoddardOlive81}P. Goddard and D.I. Olive, Nucl. Phys. B191 (1981) 511.
\bibitem{JaffeTaubes}A Jaffe and C. Taubes, Vortices and Monopoles, Birkhäuser Boston, 1980.
\bibitem{VaYung}A. Hanany, M. J. Strassler and A. Zaffaroni, Nucl.Phys. \textbf{B513} (1998)
87; J. D. Edelstein, W. G. Fuertes, J. Mas, J. M. Guilarte, Phys.Rev. \textbf{D62}
(2000) 65008; A. Vainshtein and A. Yung, hep-th/0012250.
\bibitem{Sohnuis}M. Sohnuis, Phys. Rept. \textbf{128} (1985) 39. 
\end{thebibliography}
\end{document}